\documentclass[aps,reprint,twocolumn,preprintnumbers,floatfix,nofootinbib]{revtex4-1}
\usepackage[utf8]{inputenc}
\usepackage[dvips]{graphicx}
\usepackage{xcolor}
\usepackage{color}
\usepackage{relsize}
\usepackage{graphics}
\usepackage{epstopdf}
\usepackage{hyperref}
\usepackage{mathrsfs}
\usepackage{ragged2e}
\usepackage{amssymb}
\usepackage{placeins}
\usepackage[bottom]{footmisc}

\usepackage[normalem]{ ulem }
\usepackage{amsthm}
\usepackage{amsmath}
\usepackage{hyperref}
\usepackage{cancel}
\usepackage{comment}

\newcommand{\abs}[1]{\left| #1 \right|}
\newcommand{\beq}{\begin{equation}}
\newcommand{\eeq}{\end{equation}}
\newcommand{\beqn}{\begin{eqnarray}}
\newcommand{\eeqn}{\end{eqnarray}}
\newcommand{\nn}{\nonumber}

\def\ie{{\it i.e.},\/}
\def\eg{{\it e.g.},\/}

\newcommand{\rhogw}{\rho_{\rm GW}}
\newcommand{\Omegagw}{\Omega_{\rm GW}}
\newcommand{\Omegasw}{\Omega_{\rm sw}}
\newcommand{\Omegaref}{\Omega_{\rm ref}}
\newcommand{\rhocrit}{\rho_{\rm crit}}
\newcommand{\fpk}{f_{\rm pk}}
\newcommand{\fref}{f_{\rm ref}}

\begin{document}
\sloppy

\vspace*{1mm}

\title{LIGO as a probe of Dark Sectors}
\preprint{UCI-HEP-TR-2021-06}

\author{Fei Huang$^{a,b}$}
\email{huangf4@uci.edu}
\author{Veronica Sanz$^{c,d}$}
\email{veronica.sanz@uv.es}
\author{Jing Shu$^{a,e,f,g,h,i}$}
\email{jshu@itp.ac.cn}
\author{Xiao Xue$^{a,e}$}
\email{xuexiao@itp.ac.cn}

\affiliation{$^a$ CAS Key Laboratory of Theoretical Physics, Institute of Theoretical Physics,
Chinese Academy of Sciences, Beijing 100190, China}
\affiliation{$^b$ Department of Physics and Astronomy, University of California, Irvine, CA 92697 USA}
\affiliation{$^c$ Instituto de F\'isica Corpuscular (IFIC), Universidad de Valencia-CSIC, E-46980 Valencia, Spain}
\affiliation{$^d$ Department of Physics and Astronomy, University of Sussex, Brighton BN1 9QH, UK}
\affiliation{$^e$ School of Physical Sciences, University of Chinese Academy of Sciences, Beijing 100049, P. R. China}
\affiliation{$^f$ CAS Center for Excellence in Particle Physics, Beijing 100049, China}
\affiliation{$^g$ Center for High Energy Physics, Peking University, Beijing 100871, China}
\affiliation{$^h$ School of Fundamental Physics and Mathematical Sciences,
Hangzhou Institute for Advanced Study, UCAS, Hangzhou 310024, China}
\affiliation{$^i$
International Centre for Theoretical Physics Asia-Pacific, Beijing/Hangzhou, China}

\begin{abstract}
We show how current LIGO data is able to probe interesting theories beyond the Standard Model, particularly Dark Sectors where a Dark Higgs triggers symmetry breaking via a first-order phase transition. We use publicly available LIGO O2 data to illustrate how these sectors, even if disconnected from the Standard Model, can be probed by  Gravitational Wave detectors. We link the LIGO measurements with the model content and mass scale of the Dark Sector, finding that current O2 data is testing  a broad set of scenarios where the breaking of $SU(N)$ theories with $N_f$ fermions is triggered by a Dark Higgs at scales $\Lambda \simeq 10^8 - 10^9$ GeV with reasonable parameters for the scalar potential.  
\end{abstract}

\maketitle

\section{Introduction}

Much of the Universe is dark, and many theories have been built trying to explain it. 
Our hopes for probing these theories often rely on their possible connection to regular matter via some form of non-gravitational interaction 
For example, direct searches for Dark Matter hinges on some sort of coupling to nucleons or electrons, and constraints on those couplings usually assume a mechanism of communication between the Dark Sector and the rest of the Universe which establishes some form of tracking between these two sectors.

The first observation of Gravitational Waves (GW) by the LIGO and Virgo collaborations~\cite{Abbott:2016blz,aasi2015advanced,acernese2014advanced} in 2015 initiated a new way to see the Universe, and since then exciting new observations have provided information about astrophysical objects like Black Holes~\cite{abbott2019gwtc,abbott2020gwtc}. However, the physics reach for LIGO is not circumscribed to detection of mergers \cite{Allen:1997ad,maggiore2000gravitational,regimbau2011astrophysical,Caprini:2018mtu}. The detection, or the lack of, a stochastic GW background allows us to explore interesting, non-standard sectors.  
We will explain how, with the current public data from LIGO, one can probe plausible Dark Sector scenarios, regardless of their non-gravitational interaction with visible matter.

These Dark Sectors could resemble Standard-Model dynamics, with new forces, Dark Higgses and states charged under them. Influenced by thermal contributions from the degrees of freedom in the Dark Sector, the thermal history of the Dark Higgs could then lead to first-order phase transitions. 
Many studies have been devoted to the prospects that future interferometers could offer to explore Dark Sectors, \eg~Ref~\cite{Croon:2018erz}. 
In this paper we explore the possibilities that LIGO and its current public dataset present, and bridge the gap between generic studies of thermal parameters, \eg~Ref~\cite{Romero:2021kby}, and specific particle-physics models.  

 The paper is structured as follows. In Sec.~\ref{secGWs} we first describe the analysis of GWs from first-order phase transitions, then discuss in Sec.~\ref{secUV} the connection between the phase transition thermal parameters with classes of particle-physics models, especially of $SU(N)/SU(N-1)$ Dark Sectors. 
 We finally link these models to current exclusions set by LIGO, and in Sec.~\ref{secConcl} we conclude the discussion.


\section{Gravitational Waves from Phase Transitions}~\label{secGWs}

The stochastic gravitational wave background (SGWB) is often considered as an isotropic, unpolarized, stationary and Gaussian background generated by a large number of unresolved gravitational-wave sources~\cite{Allen:1997ad,maggiore2000gravitational,regimbau2011astrophysical,Caprini:2018mtu}.
Its power spectrum is characterized by the dimensionless quantity
\beq
\Omegagw(f)\equiv \frac{1}{\rhocrit}\frac{d\rhogw}{d\ln f},
\eeq
where $\rhogw$ is the energy density of the stochastic gravitational wave background, $f$ is the frequency of the GW and
\beq
\rhocrit\equiv \frac{3c^2H_0^2}{8\pi G}
\eeq
is the critical energy density of the universe today.

In principle, the total SGWB is a superposition of all possible astrophysical and cosmological sources.
However, we can obtain a conservative upper limit for the SGWB from phase transitions that occur in the early universe by assuming that phase transition dynamics is the main source of SGWB.

\begin{figure}[t]\centering
\includegraphics[width=0.45\textwidth]{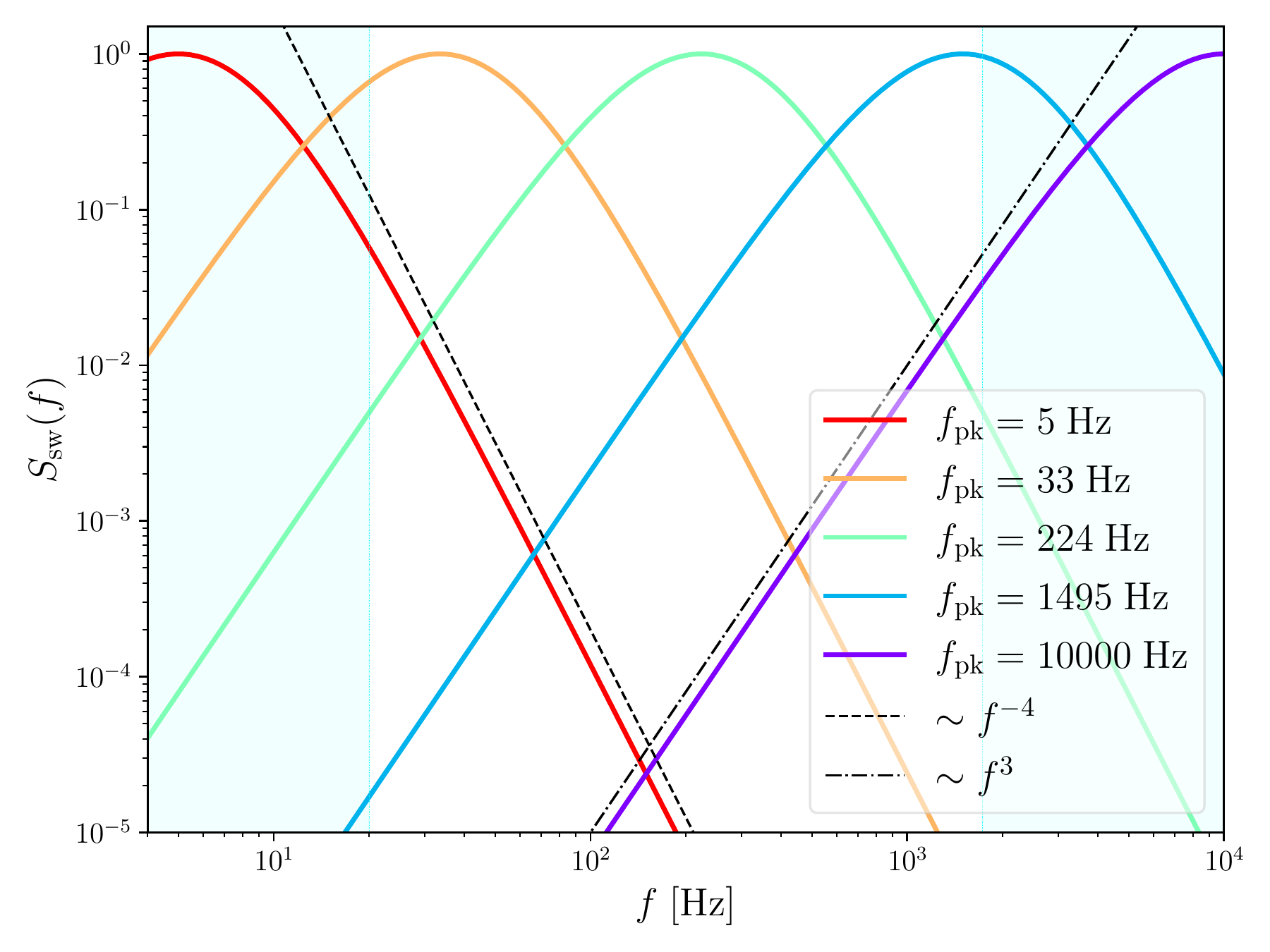}
\caption{The coloured curves show the spectral shape $S_{\rm sw}(f)$ by varying the peak frequency $f_{\rm sw}$, whereas the dashed and dotted black lines indicate its asymptotic behaviour. 
The two solid vertical lines at $20$ and $1726~\rm Hz$  show the minimum and maximum frequency considered when obtaining the GW upper limits using LIGO data.
}\label{fg:spectral_shape}
\end{figure}

The SGWB generated from phase transitions in the early universe consists of three parts~\cite{Caprini:2007xq,Huber:2008hg,Caprini:2009yp,Espinosa:2010hh,Hindmarsh:2013xza,Hindmarsh:2016lnk}:
\beq
\Omegagw=\Omega_{\rm col}+\Omegasw+\Omega_{\rm turb}\,,
\eeq
in which the three terms on the right hand side correspond to the contribution from bubble collisions, sound waves in the fluid and the turbulence, respectively.
For simplicity, we shall assume in this work that contributions from sound waves are always dominant. We emphasize that this is typically the case for models in which gauge bosons acquire masses during the phase transition \cite{Bodeker:2017cim}.
However, the analysis we present can be easily generalized to cases in which other types of contribution become more important.

The phase transition is in general characterized by just a few parameters: the velocity of the bubble wall $v_w$, the ratio of the free energy density difference between the true and false vacuum and
the total energy density, $\xi$, the speed of the phase transition $\beta/H$, and the nucleation temperature $T_N$.
With these parameters, the GW power spectrum can be expressed as \cite{Weir:2017wfa}
\beq
\Omegagw h^2 \simeq 8.5\times 10^{-6}\left(\frac{g_*}{100}\right)^{1/3}\Gamma^2\bar{U}_f^4\left(\frac{\beta}{H}\right)^{-1}v_wS_{\rm sw}(f), \label{eq:gw_sw}
\eeq
where $g_*$ is the effective number of relativistic degrees of freedom at the time of the transition, $\Gamma\sim 4/3$ is the adiabatic index, $\bar{U}_f^4\sim (3/4)\kappa_f\xi$ is the root-mean-square fluid velocity with the efficiency parameter given by the approximate expressions~\cite{Espinosa:2010hh}
\beq
\kappa_f\sim 
\begin{cases}
\displaystyle\frac{\xi}{0.73+0.83\sqrt{\xi}+\xi} & v_w\rightarrow 1\\
\displaystyle\frac{\xi^{2/5}}{0.017+(0.997+\xi)^{2/5}} & v_w \approx 0.5
\end{cases}\,,
\eeq
The spectral shape $S_{\rm sw}$ is given by
\beq
S_{\rm sw}(f)=\left(\frac{f}{f_{\rm pk}} \right)^3 \left( \frac{7}{4+3\left(f/f_{\rm pk}\right)^2}\right)^{7/2}\label{eq:Ssw}
\eeq
with the peak frequency
\beq
f_{\rm pk} = 8.9\times 10^{-8} \mathrm{Hz} \left(\frac{1}{v_w}\right) \left(\frac{\beta}{H}\right) \left(\frac{T_N}{\rm GeV}\right) \left(\frac{g_*}{100}\right)^{1/6}.
\eeq
The shape of $S_{\rm sw}$ is shown in FIG.~\ref{fg:spectral_shape}, noting that $S_{\rm sw}$ is equal to $\Omegagw (f)/\Omegagw (f_{pk})$. 
In this figure one observes that varying the peak frequency $f_{\rm sw}$ amounts to simply shifting the spectrum horizontally. Also note that the asymptotic behavior of
$S_{\rm sw}$ goes as $\sim f^{3}$ for $f\ll f_{\rm sw}$ (dotted line), whereas one expects a behaviour $\sim f^{-4}$ for $f\gg f_{\rm sw}$ (dashed line). 

The behavior of $S_{\rm sw}$ in the LIGO frequency range (indicated by the two vertical lines) can therefore transition from a simple descending power law, to one with a peak in between, and eventually to a ascending power law as we increase $\fpk$.

\begin{figure}[t]\centering
\includegraphics[width=0.45\textwidth]{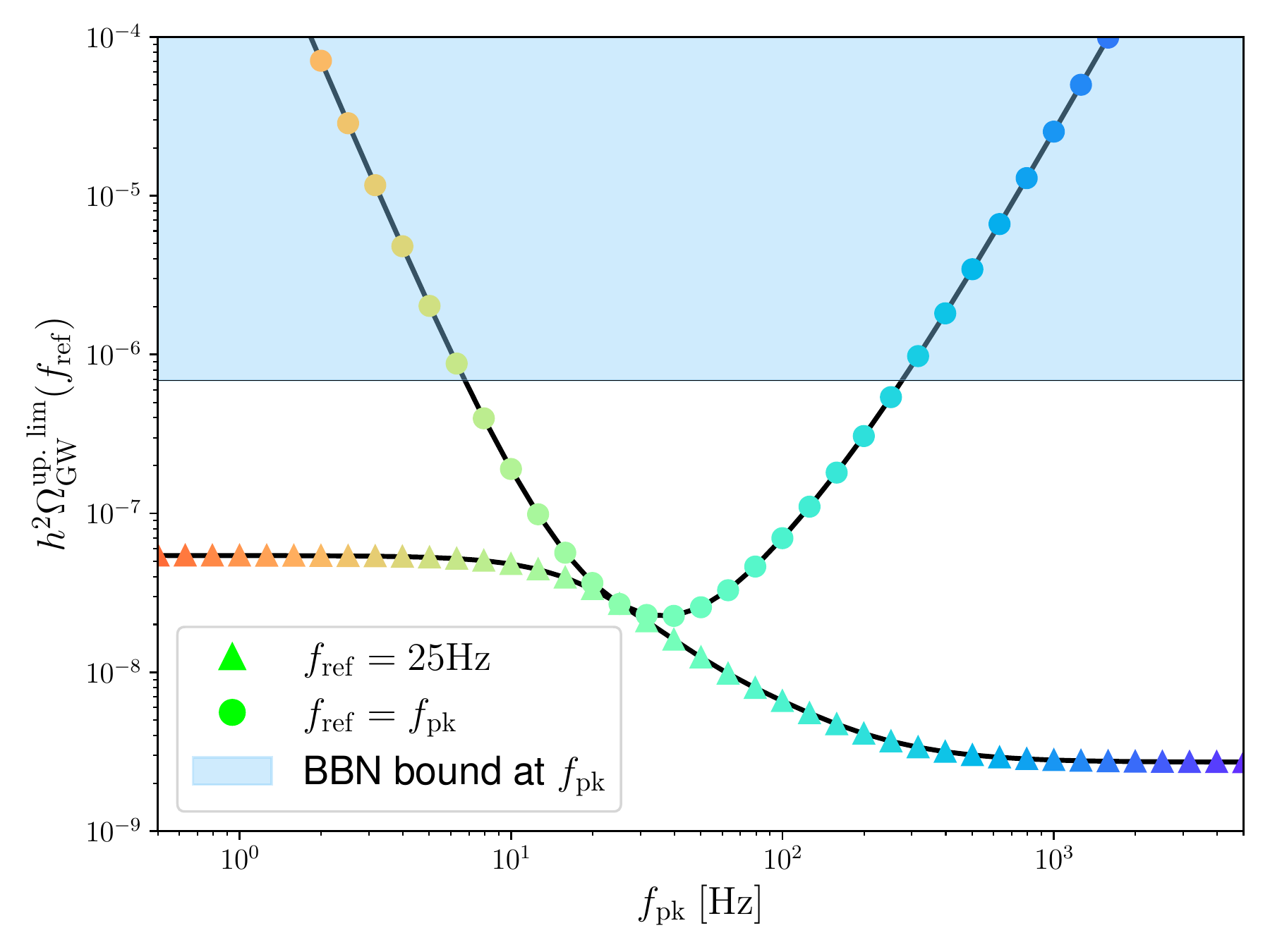}
\caption{The upper limit of $\Omegaref$ for each different $\fpk$ using the $2\sigma$ criterion. 
The curve with triangles corresponds to the upper limit at 25 Hz, while the curve with circles uses the upper limit at $\fpk$.
The solid horizontal line corresponds to the BBN bound at $\fpk$.
}\label{fg:shape_limit}
\end{figure}

With this spectrum, we follow the procedure laid out in Refs~\cite{TheLIGOScientific:2016dpb,LIGOScientific:2019vic} to compute the upper limit of $\Omegagw$ for different values of $\fpk$ using data from LIGO O2 \cite{LIGOScientific:2019vic}. More details can be found in Appendix~\ref{appA}.
The only difference respect to these references is related to the choice of the optimized estimator, \ie~ the estimator $\hat{\Omega}_{\fref}$ is optimized by putting $\Omegagw(f)/\Omegagw(\fref)$ instead of some power of $f/\fref$.
Note that this means the upper limit essentially depends only on the shape of $S_{\rm sw}$ within the LIGO frequency band, since all the other factors drop out when taking the ratio.
The 95\% confidence level upper limit is then obtained by setting
\beq
\Omegagw^{\rm up.~lim.}(\fref)=2\sigma(\fref)\,.
\eeq
In FIG.~\ref{fg:shape_limit} we show the upper limit at $\fref=25$ Hz by the curve with triangles.
For $\fpk \lesssim 10$ Hz or $\fpk \gtrsim 10^3$ Hz, the bound approaches a constant since the spectrum is essentially a power law with fixed exponent within the LIGO frequency band.
This behaviour allows us to easily extrapolate the constraint to even smaller or bigger values of $\fpk$.
However, for intermediate $\fpk$ there is a smooth transition on the upper limit between the ascending and descending asymptotic behavior.
We see that LIGO provides a stronger constraint on ascending spectra than on the descending spectra, and the difference can be as large as an order of magnitude.

In practice, LIGO could only provide reliable constraints on SGWB in the band $20-1726$ Hz and we have chosen $\fref=25$ Hz as it is the frequency where LIGO is most sensitive~\cite{TheLIGOScientific:2016dpb,LIGOScientific:2019vic}.
However, the thermal parameters of the phase transition are connected to the amplitude at the $\fpk$.
In order to place constraint on the thermal parameters, we notice that, for a given GW spectrum as in Eq.~(\ref{eq:Ssw}), the constraint at a particular frequency in the LIGO band can be mapped to the peak frequency $\fpk$ using
\beq
\Omegagw^{\rm up.~lim.}(\fpk)=\Omegagw^{\rm up. ~lim}(\fref)\frac{S_{\rm sw}(\fpk)}{S_{\rm sw}(\fref)}\,.
\eeq
The curve made by circles in FIG.~\ref{fg:shape_limit} shows the result of this mapping.
For $\fpk$ very small or very large, the upper bound from LIGO becomes substantially weaker, even when comparing with the constraint from BBN, which is obtained by integrating out the spectrum and requiring $h^2\int d\ln f~ \Omegagw(f)<1\times 10^{-6}$ \cite{Caprini:2018mtu,Tanin:2020qjw}.

In what follows, we shall discuss the LIGO constraints on the thermal parameters and how such constraint can be utilized to constrain particular models of phase transition.

%

\section{Scenarios for phase transitions and their thermal parameters}~\label{secUV}

Typically, one represents phase transitions as driven by the dynamics of a scalar field which transitions from one vacuum to another under the influence of the evolving thermal potential.
In that context, the thermal parameters $\beta/H$ and $\xi$ are defined by
\beqn
\frac{\beta}{H}&=&\left.T\frac{(S_E/T)}{dT}\right|_{T=T_N}\\
\xi&=&\frac{1}{\rho_N}\left.\left( \Delta V -T\Delta\frac{dV}{dT} \right)\right|_{T=T_N}
\eeqn
in which $S_E$ is the Euclidean action, $V$ is the thermal potential of the scalar field, 
and the bubble nucleation temperature can be obtained by solving
\beq
\frac{S_E}{T_N} \approx 177 -4 \ln \left(\frac{T_N}{\rm GeV}\right) -2\ln g_{\star}(T_N)\,,\\
\eeq
in which $\rho_N=g_\star(T_N)\pi^2T_N^4/30$.
On the other hand, the calculation of the bubble-wall velocity $v_w$ for a particular model is highly non-trivial.
Therefore, instead of directly calculating it, we shall follow the customary convention of considering a few reference values, $v_w=0.5$ and $v_w=1$.

To connect the thermal parameters, $\xi$ and $\beta/H$ to specific models we will follow the approach described in Ref.~\cite{Croon:2018erz}. We will consider classes of potentials which consists of competing terms with alternating signs.
Specifically, we will look into two types of finite-temperature potentials
\beqn
V(H,T)&=& \frac{1}{2}m^2(T)h_D^2-c_3(T)h_D^3+\frac{1}{4}\lambda(T)h_D^4\label{eq:finite_temp_1}\\
V(H,T)&=& \frac{1}{2}m^2(T)h_D^2-\frac{1}{4}\lambda(T)h_D^4+c_6(T)h_D^6
\label{eq:finite_temp_2}
\eeqn
in which the coefficients of the scalar field $h_D$ are all positive at the time of transition.

Indeed, most dark phase transitions can be mapped onto these effective scenarios.
For example, in a Dark Sector where its particles acquire mass from a Dark Higgs as its gauge group $SU(N)$ breaks into $SU(N-1)$, 
the potential in Eq.~(\ref{eq:finite_temp_1}) can be realized with renormalisable operators, whereas the potential in Eq.~(\ref{eq:finite_temp_2}) can be obtained from non-renormalisable sextet interaction \cite{Croon:2018erz}.

In what follows, we shall discuss the Dark Higgs - $SU(N)/SU(N-1)$ models with renormalisable and non-renormalisable operators specifically.

\subsection{Exploring Dark Sectors with LIGO: \texorpdfstring{$SU(N)/SU(N-1)$}{} models} 

\subsubsection{Models with renormalisable operators}
For the type of potential in Eq.~(\ref{eq:finite_temp_1}),
we can parametrise  zero-temperature parameters as
\beq
m^2(0)=-\frac{\Lambda^4}{v_D^2},\,\,\,\lambda(0)=\frac{\Lambda^4}{v_D^4}
\eeq
in which $v_D$ is the zero temperature vacuum expectation value, and $\Lambda$ is the scale of the potential.
With this parametrisation, the finite temperature potential can be expressed as
\begin{widetext}
\beqn
V(H,T)&=&\Lambda^4\left[-\frac{1}{2}\left(\frac{h_D}{v_D}\right)^2+\frac{1}{4}\left(\frac{h_D}{v_D}\right)^4\right] + \frac{T^4}{2\pi^2}\left[ \sum_{i\in \rm bosons}n_i J_B(m_i^2/T^2) - \sum_{i\in \rm fermions}n_i J_F(m_i^2/T^2)\right]\nn\\
&=&\Lambda^4\Bigg\{ \left[-\frac 1 2+\left( \frac 1 8 + \frac{N_G}{24} \right)\frac{T^2}{v_D^2} +\frac{3}{24} N_{\rm GB}\frac{g^2}{4}\frac{T^2v_D^2}{\Lambda^4} +y^2N_f\frac{T^2}{48}\frac{v_D^2}{\Lambda^4} \right]\left(\frac{h_D}{v_D}\right)^2\nn\\
&~&~~~~~~~-\left[ N_{\rm GB} \left(\frac{g^2}{4}\right)^{3/2}\frac{1}{4\pi} \frac{v_D^3T}{\Lambda^4} \right]\left(\frac{h_D}{v_D}\right)^3+\frac 1 4 \left(\frac{h_D}{v_D}\right)^4\Bigg\}\,,\label{eq:re_h_T_expansion}
\eeqn
\end{widetext}
where $N_{\rm GB}=2N-1$ is the number of gauge bosons that couple to the Dark Higgs with coupling constant $g$, and which get a mass from the Dark Higgs interactions. $N_{G}=2N-1$ is the number of Goldstone bosons, and \mbox{$N_f=N\times N_{\rm FL}$} is the number of self-adjoint  fermions with Yukawa coupling $y$ where $N_{\rm FL}$ is the number of flavors. Note that for Dirac fermions, one would need to double the number of degrees of freedom.
For simplicity, we assume the Yukawa coupling is universal for those fermions. See Ref.~\cite{Croon:2018erz} for more details.

In the second equality, the following high temperature expansions are used
\begin{eqnarray}
J_B(m^2/T^2)\sim 2\pi^2\left(\frac{m^2}{24T^2}-\frac{m^3}{12\pi T^3}\right),\\
~~\text{and}~~J_F(m^2/T^2) \sim -2\pi^2\left(\frac{m^2}{48T^2}\right),
\end{eqnarray}
in which the field dependent masses can be read as
\beqn
m_H^2&=&\partial_{h_D}^2 V =\Lambda^4\left( 3\frac{h_D^2}{v^4}-\frac{1}{v^2}\right)\,,\\
m_G^2&=&\frac{1}{h_D}\partial_{h_D} V=\Lambda^4\left(\frac{h_D^2}{v_D^4}-\frac{1}{v_D^2}\right)\,,\\
m_{\rm GB}&=&\frac{gh_D}{2}\,,\\
m_f&=&\frac{yh_D}{\sqrt{2}}.
\eeqn
Note that in the second line of Eq.~(\ref{eq:re_h_T_expansion}), only the massive gauge bosons are taken into account, \ie~the higher order term ($\sim m^3/T^3$) from the Goldstone boson is neglected.Besides, the part of the expansion which gives rise to terms independent of $h_D$ is neglected, since it only amounts to a constant shift in the potential $V$. Finally, the mapping to the temperature dependent parameters in Eq.~(\ref{eq:finite_temp_1}) is straightforward by matching the terms with the same powers of $h_D$.

Eq.~(\ref{eq:re_h_T_expansion}) is also often written in the following form
\beqn
V(H,T)=\Lambda^4(T)\Bigg[& & \left(\frac{3-4\alpha(T)}{2}\right)\left(\frac{h_D}{v_D(T)}\right)^2 \nn\\
&~&- \left(\frac{h_D}{v_D(T)}\right)^3+ \alpha(T)\left(\frac{h_D}{v_D(T)}\right)^4 
\Bigg]\,,\nn\\
\eeqn
in which the minimum of the potential can be easily obtained by minimizing the potential
\beq
v_D(T)=\frac{3c_3(T)+\sqrt{9c_3^2(T)-4m^2(T)\lambda^2(T)}}{2\lambda(T)}.
\eeq
By identifying 
\beqn
c_3(T)=\frac{\Lambda^4(T)}{v_D^3(T)},~~~~~
\frac{\lambda(T)}{4}=\alpha(T)\frac{\Lambda^4(T)}{v_D^4(T)}\,,
\eeqn
one finds
\beqn
\alpha(T)=\frac{\lambda(T)v_D(T)}{4c_3(T)},~
\Lambda(T)=\left( c_3(T)v_D^3(T) \right)^{1/4}\,.
\eeqn
With these, for $\alpha(T) \in
[0.51, 0.65]$, the effective action can be fitted by \cite{Croon:2018new}
\beq
\frac{S_E}{T}=\frac{v_D^3(T)}{T\Lambda^2(T)}10^{a+b\frac{\abs{\alpha(T)-0.75}^c}{\abs{\alpha(T)-0.5}^d}}
\eeq
with the fitting parameters \mbox{$a=-71.06,~b=71.62$}, \mbox{$c=0.008805$}, and \mbox{$d = 0.009263$}.

\subsubsection{Models with non-renormalisable operators}

For the type of potential in Eq.~(\ref{eq:finite_temp_2}), similar to the previous case, one can perform the following parametrisation
\beq ~\label{potnonren}
m^2(0)=(2-3\alpha)\frac{\Lambda^4}{v_D^2},\,\,\,\lambda(0)=4\frac{\Lambda^4}{v_D^4},\,\,\,c_6(0)=\alpha\frac{\Lambda^4}{v_D^6}.
\eeq 
The finite temperature potential then becomes
\begin{widetext}
\beqn
V(H,T)&=&\Lambda^4\left[(2-3\alpha)\left(\frac{h_D}{v_D}\right)^2-\left(\frac{h_D}{v_D}\right)^4 + \alpha\left(\frac{h_D}{v_D}\right)^6\right]\nn\\
&~&+ \frac{T^4}{2\pi^2}\left[ \sum_{i\in \rm bosons}n_i J_B(m_i^2/T^2) - \sum_{i\in \rm fermions}n_i J_F(m_i^2/T^2)\right]\nn\\
&=&\Lambda^4\Bigg\{ \bigg[2-3\alpha-\left( \frac 1 2 + \frac{N_G}{6} \right)\frac{T^2}{v_D^2} +\frac{3}{24} N_{\rm GB}\frac{g^2}{4}\frac{T^2v_D^2}{\Lambda^4} +y^2N_f\frac{T^2}{48}\frac{v_D^2}{\Lambda^4} \bigg]
 \left(\frac{h_D}{v_D}\right)^2\nn\\
&~&~~~~~-\left[1-\frac{(30+6N_G)\alpha T^2}{24v_D^2} \right]\left(\frac{h_D}{v_D}\right)^4 +\alpha \left(\frac{h_D}{v_D}\right)^6\Bigg\}\label{eq:nonre_h_T_expansion}\,.
\eeqn
\end{widetext}
Note that, in the high-temperature expansion, the cubic term is assumed to be subdominant, \ie~we have only kept the part proportional to $m^2/T^2$ \cite{Croon:2018erz}.
Moreover, the field-dependent masses of the Goldstone bosons and the Dark Higgs which goes into the thermal correction are
\beqn
m_H^2&=&\Lambda^4\left[ \frac{2(2-3\alpha)}{v_D^2} -\frac{12h_D^2}{v_D^4} +  \frac{30\alpha h_D^4}{v_D^6}\right]\,,\\
m_G^2&=&\Lambda^4\left[ \frac{2(2-3\alpha)}{v_D^2} -\frac{4h_D^2}{v_D^4} +  \frac{6\alpha h_D^4}{v_D^6}\right]\,.
\eeqn
The terms proportional to $\alpha h_D^4/v_D^4$ would give rise to the thermal correction of the quartic term.

Just as we have done in the previous section, one can write Eq.~(\ref{eq:nonre_h_T_expansion}) in terms of temperature-dependent parameters:
\beqn
V(H,T)=\Lambda^4(T) \Bigg[& &(2-3\alpha(T)) \left(\frac{h}{v_D(T)}\right)^2 
-\left(\frac{h}{v_D(T)}\right)^4\nn\\
& &+\alpha(T)\left(\frac{h}{v_D(T)}\right)^6
\Bigg]\,.
\eeqn
The non-vanishing VEV 
\beq
v_D(T)=\left(\frac{\lambda(T)+\sqrt{\Lambda^2(T)-24c_6(T)m^2(T)}}{12c_6(T)}\right)^{1/2}
\eeq
is obtained by minimizing the potential.
Suppose the non-vanishing VEV does exists ($\Lambda^2(T)-24c_6(T)m^2(T)>0$), then $\alpha(T)$ can be obtained by solving 
\beq
\lambda(T)=4\Lambda(T)^4/v(T)^4\,\,\,c_6(T)=\alpha(T)\Lambda(T)^4/v_D(T)^6.
\eeq
Therefore,
\beq
\alpha(T)=4c_6(T)v_D^2(T)/\lambda(T).
\eeq
Following \cite{Croon:2018erz}, the Euclidean action can be fitted by
\beq
S_E=\frac{v_D^3(T)}{\Lambda^2(T)} 10^{\sum_{i=1}^3 a_i(\alpha(T)-2/3)^i}
\eeq
with $a_i=(-17.446, -132.404, -763.744)$ for $\alpha(T)\in [0.51,0.65]$.

Finally, for $g_\star (T_N)$,
at high temperatures, \ie~ {$T\gg\mathcal{O}(100)~\text{GeV} \gg m_i,~m_{h_D}$}, we shall assume that the particles in the Dark Sector are the only degrees of freedom in addition to the SM. 
Therefore, the effective number of relativistic degrees of freedom will be given by
\beq 
g_\star(T_N)\approx 106.75+3N_{\rm GB}+2(N^2-1-N_{\rm GB})+1+\frac{7}{8}\times 2 \times N_f\,,
\eeq
where we have included all degrees of freedom from the SM, as well as the massless and massive gauge bosons and the dark fermions charged under $SU(N)$.
\subsubsection{Results}

\begin{figure*}[t]
\centering
\includegraphics[width=0.49\textwidth]{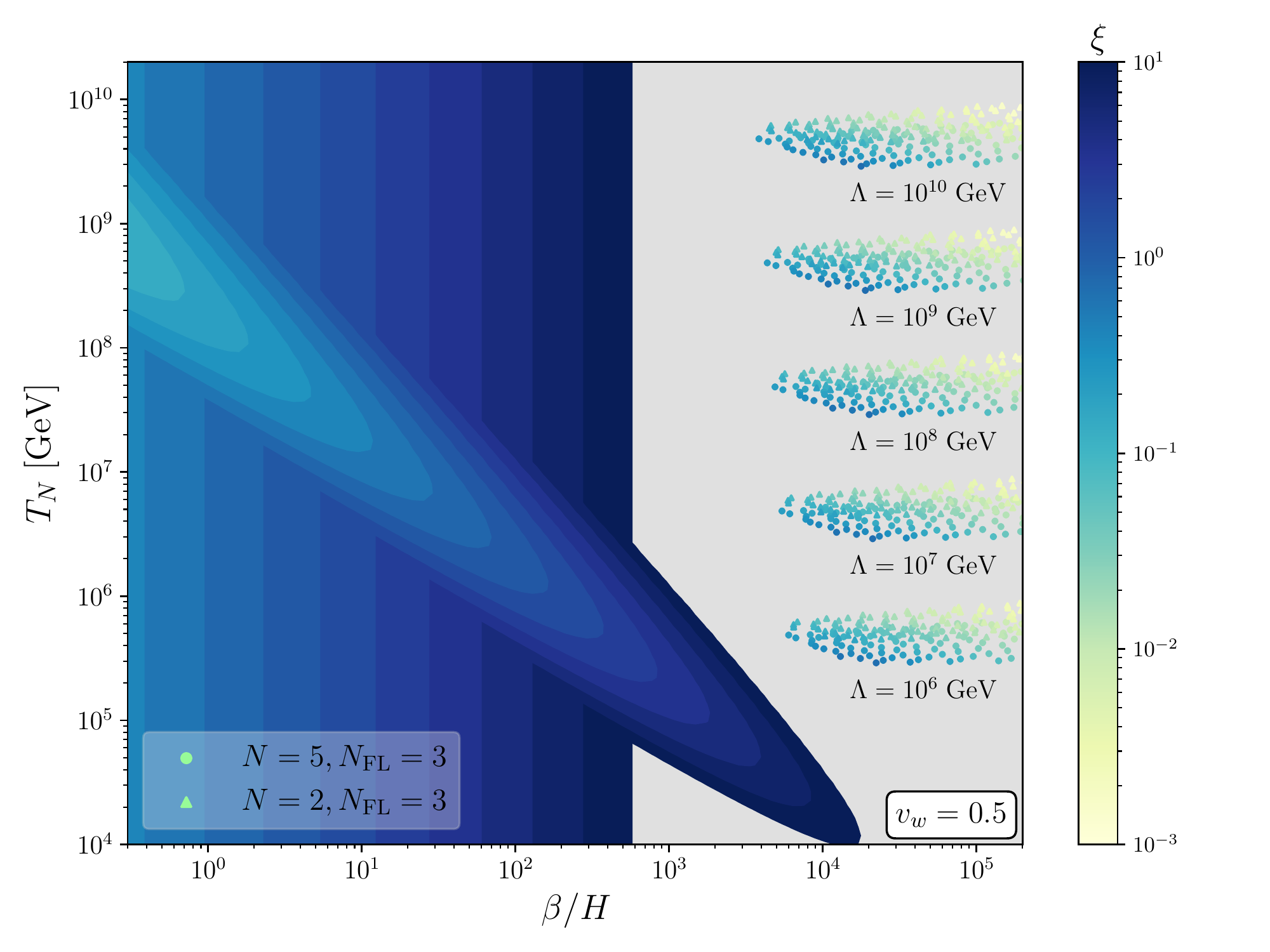}\includegraphics[width=0.49\textwidth]{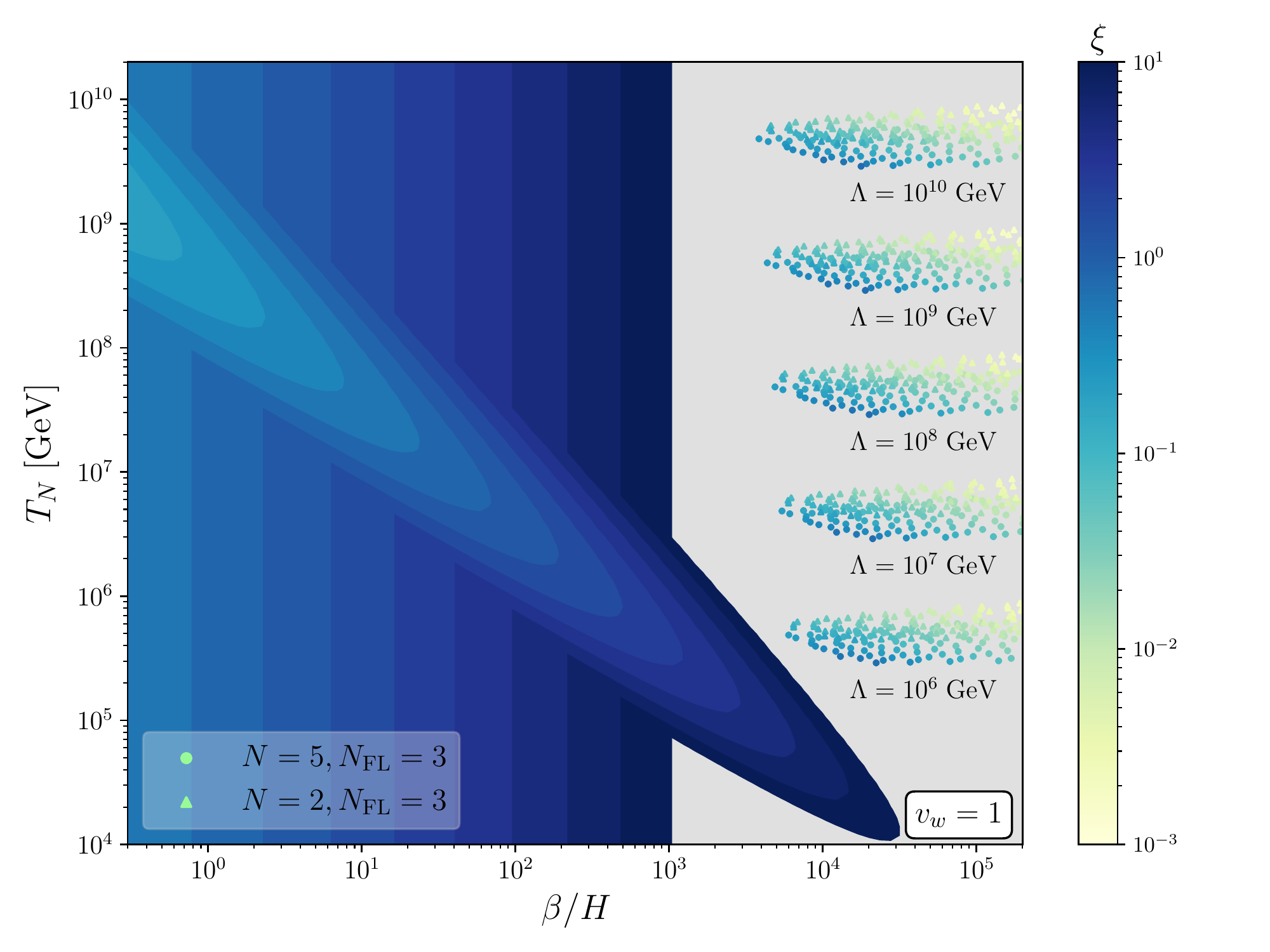}
\includegraphics[width=0.49\textwidth]{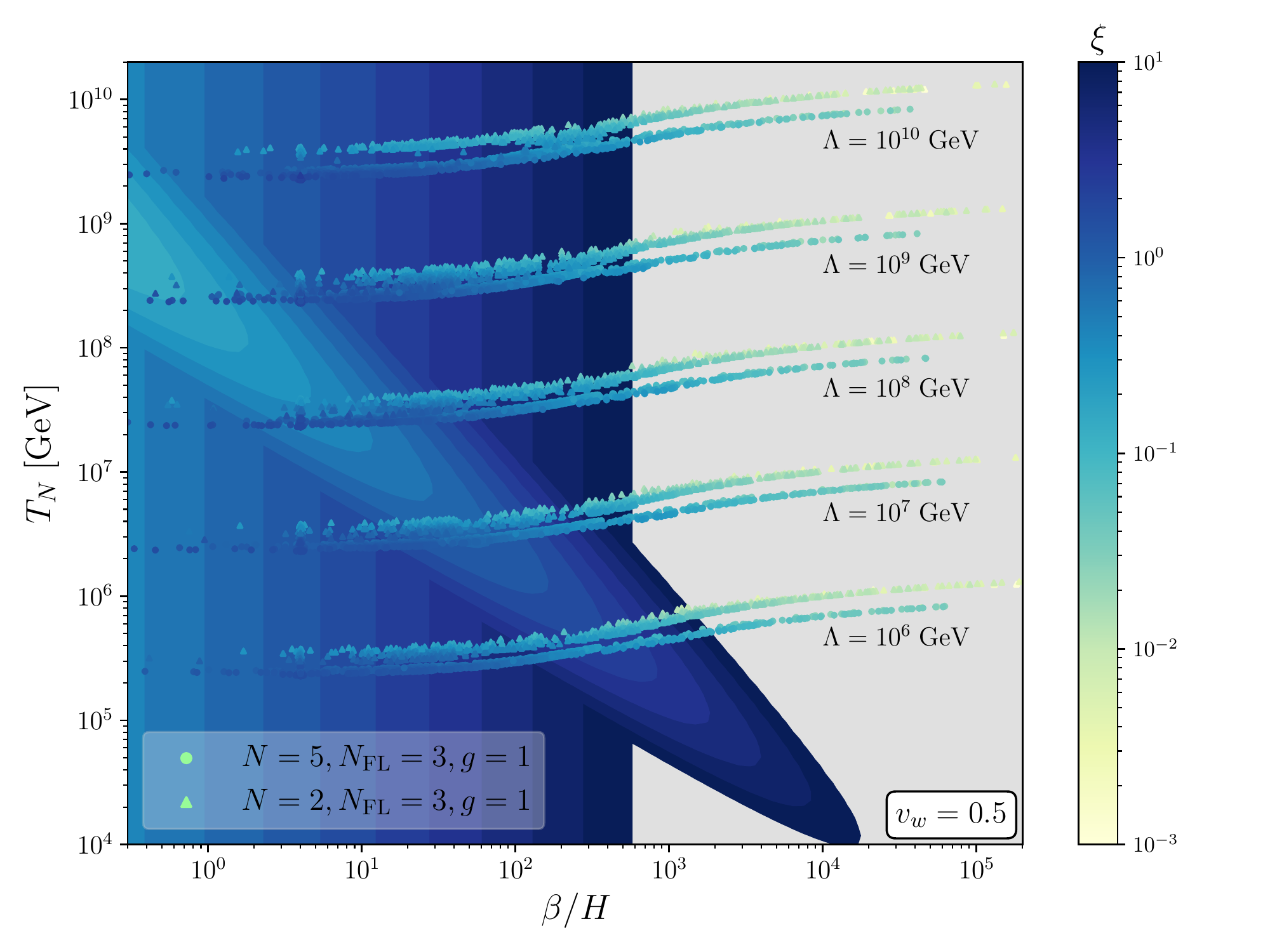}\includegraphics[width=0.49\textwidth]{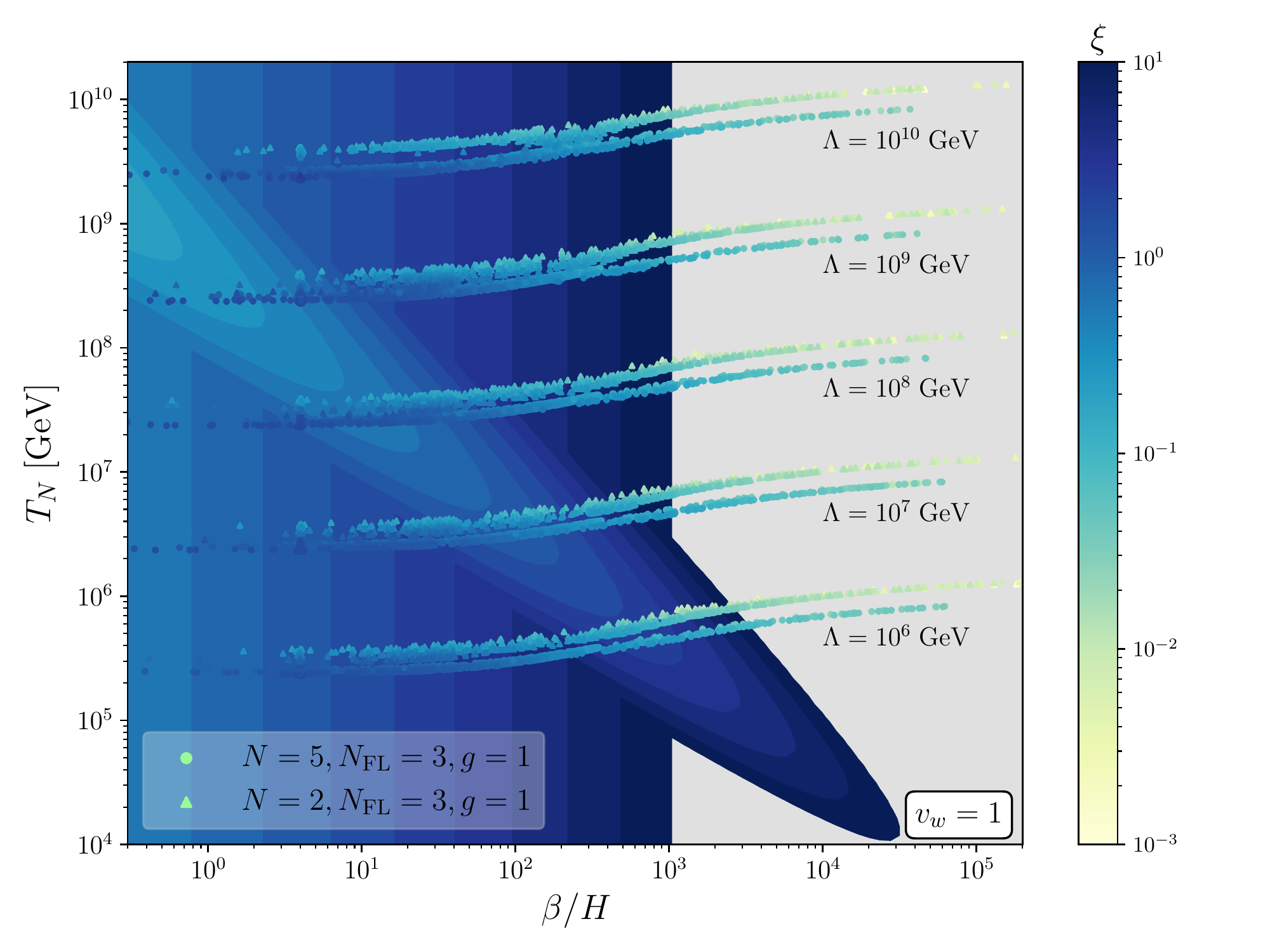}
\caption{95\% confidence level exclusion region from LIGO (curved contours) and the constraint from BBN (vertical contours), projected on the $T_N-\beta/H$ plane. The values of $\xi$ corresponding to this exclusion are represented by the colour variation.
The constraints are calculated for two reference bubble-wall velocity $v_w=0.5$ and $v_w=1$.
Markers with different shapes are obtained from different models of phase transitions. 
The top two panels correspond to the cases with renormalisable operators, whereas the bottom two panels show the cases with non-renormalisable operators.
}\label{fg:2D}
\end{figure*}   

In FIG.~\ref{fg:2D}, we present the LIGO constraints on the thermal parameters at 95\% confidence level together with the constraint from BBN.
In the left panels, the constraints are evaluated with $v_w=0.5$, whereas the bounds in the right panels are evaluated at $v_w=1$.

Note that, with $v_w$ fixed, the thermal parameters $(\beta/H, T_N, \xi)$ constitutes a 3-dimensional space.
In our analysis, we find the the excluded region can be conveniently projected on the 2-dimensional plane of $\beta/H$ and $T_N$ as the contours of $\xi$ do not intersect each other. 
This thus enables us to use colour variation to represent the values of $\xi$ on the exclusion contours.
Indeed, within a particular contour, any value of $\xi$ larger than the value associated with the contour is excluded.
As a result, 
the regions in the 2D plots where the LIGO ellipsoidal contours have a colour lighter than the BBN vertical contours suggest that LIGO can set a better bound than BBN, and viceversa.

On the same figure, we also plot the values of the thermal parameters obtained from different models defined by $\Lambda,~N$ and $N_{\rm FL}$. We show examples with $N=2$ and 5, and a fixed number of flavours $N_{\rm FL}=3$, as representative of the behaviour expected from the choices on matter content and gauge groups. We also fix $g=1$ and $y=1$, although similar behaviour is found for similar ${\cal O}(1)$ values.

If a set of thermal parameters obtained from a particular model lies within a contour, and has a larger $\xi$ (darker colour) than the corresponding contour, then this set of thermal parameter is excluded at 95\% CL.

From the top figures~\ref{fg:2D}, one can deduce that the constraint from LIGO is still unable to probe the renormalisable classes of models, as the markers representing the thermal parameters barely touch the excluded region.

However, in the non-renormalisable case, we indeed see how LIGO is able to probe these Dark Sectors. In particular we observe that the dark scales $\Lambda \sim 10^8$ - $10^9$ GeV are best constrained by LIGO. 
Note that models with larger group rank $N$ and number of flavours $N_{\rm FL}$ are more constrained as they tend to produce a larger $\xi$.

The regions excluded by LIGO O2 correspond to $\Lambda$ in the region around $10^8$-$10^9$ GeV. In this region we have varied $v_w$ from 0.5 and 1 and explored different options for $N$ and $N_{\rm FL}$ as illustrated  by FIG.~\ref{fg:2D}. The range of parameters $\alpha$ and $x\equiv v_D/\Lambda$ where LIGO currently has sensitivity is given by 
\begin{equation}
\alpha \in [0.6,0.65] \, , \, x=\frac{v_D}{\Lambda} \in [1.4, 2] \,.
\end{equation}

These regions can be translated into parameters in the scalar potential Eq.~(\ref{eq:finite_temp_2}) by inspecting Eq.~(\ref{potnonren}):
\begin{eqnarray}
\frac{m}{\Lambda} \in [0.1,0.35],\, \lambda \in [0.2-1.1],\, c_6 \Lambda^2 \in [0.01,0.1].\, 
\end{eqnarray}
Clearly, these are reasonable choices of parameter space and indicate that LIGO is testing interesting Dark-Sector theories.

\section{Conclusions}~\label{secConcl}

When discussing Dark Sectors and their GW signatures, we usually think on  future probes like LISA, many years from now. Here we have shown that LIGO is {\it already} probing interesting scenarios for Dark Sectors.

In the context of first-order phase transitions and LIGO data, the emphasis has been placed in performing effective analyses, such as Ref.~\cite{Romero:2021kby} where the authors explore the bounds from LIGO O3 using parametrisations of the power spectrum. 
In this paper, we take a complementary step and focus on examining whether concrete particle-physics models could be related to the tested regions.

In this work we answer the question whether the type of first-order phase transitions LIGO is currently probing could be represented by concrete, reasonable particle-physics models.
For that reason, we have focused on classes of models which capture a broad set of features of Dark Sectors, and, at the same time, capable of producing interesting GW signatures.
The renormalisable and non-renormalisable benchmarks we used had been identified in Ref.~\cite{Croon:2018erz} as more promising for strong first-order phase transitions from Dark Sectors. 

We choose to set up those models with a breaking \mbox{$SU(N)/SU(N-1)$} which should be understood as an example in which some bosonic and fermionic degrees of freedom influence the thermal history of the Dark Higgs.
Such a choice also allows a simple parametrisation in terms of the group rank $N$ and the number of flavours $N_{\rm FL}$.
We find that scales around $10^8-10^9$ GeV are better probed by LIGO O2. 
We also find that the sensitive regions correspond to moderate values for $N$ and $N_{\rm FL}$, evidencing that LIGO is not testing extreme regions in the UV  parameter space.
Of course, various other types of models for phase transition could also generate GW whose spectrum lies in the frequency range relevant for LIGO, \eg~models motivated by grand unification theories \cite{Croon:2018kqn,Croon:2019kpe} and models for confinement-deconfinement phase transition \cite{Huang:2020mso}. 
It is straightforward to see whether LIGO might be able to constrain those models once the thermal parameters are computed.

Note that, when obtaining the GW peak amplitude in Eq.~(\ref{eq:gw_sw}), we have used the standard formula from Ref.~\cite{Weir:2017wfa}. 
Recent discussions in Ref.~\cite{Guo:2020grp} suggest the existence of an additional suppression factor due to the finite lifetime of the sound waves. 
Moreover, it has been shown recently that theoretical uncertainties in GW production could lead to changes in the GW spectrum as large as several orders of magnitude \cite{Croon:2020cgk}.
In addition, we have assumed in our analysis that sound waves are the dominant source for GW production.
In other scenarios, for example, strongly supercooled phase transitions \cite{Ellis:2019oqb,Ellis:2020nnr,Lewicki:2020jiv,Lewicki:2020azd}, or scenarios in which new heavy particles can provide sufficiently large friction \cite{Vanvlasselaer:2020niz},
contribution from turbulence or bubble collision could also be important and one needs to care about their effects in the GW power spectrum.
Although subject to those uncertainties, our analysis nevertheless continues to provide a concrete method to constrain Dark-Sector models with LIGO data.

Our analysis of the SGWB is based on publicly available O2 data. 
In Appendix~\ref{appA}, we have provided explanations on how to reproduce our analysis.
Recent papers from authors in the LIGO/Virgo collaboration, \eg~\cite{Romero:2021kby,Abbott:2021xxi}, make use of the O3 data.
GW constraints on particle-physics models considered in this paper are expected to improve as more data becomes available.

We believe our results motivate a more systematic study of the particle-physics scenarios that the LIGO experiment is able to test.
We emphasize again that traditional direct or indirect searches for dark particles assume that the Dark Sector interacts non-gravitationally with the Standard Model.
On the other hand, since gravity is universal, methods for probing the Dark-Sector via its gravitational effects such as structure formation (see Ref.~\cite{Dienes:2020bmn,Dienes:2021itb} and references in it for recent progress) and gravitational waves do not rely on those assumptions. 
These gravitational effects therefore offer unique opportunities to access Dark Sectors, which would otherwise be hidden from us if they lack a connection to the Standard Model.

\section*{Acknowledgements}
We would like to thank Djuna Croon for conversations at the beginning of this project. V.S. acknowledges support from the UK Science and Technology Facilities Council ST/L000504/1.
J.S. and F.H. are supported by the National Natural Science Foundation of China under Grants No. 12025507, No. 11690022, No.11947302; and is supported by the Strategic Priority Research Program and Key Research Program of Frontier Science of the Chinese Academy of Sciences under Grants No. XDB21010200, No. XDB23010000, and No. ZDBS-LY-7003.
F.H. is also supported by the National Science Foundation of China under Grants No. 12022514 and No. 11875003.
\appendix
\section{On the use of LIGO data}~\label{appA}
In this Appendix, we describe what we actually do with data downloaded from LIGO.
This consists of two parts: 1) the data selection in which we select data that satisfy certain criterion, and 2) the analysis in which we estimate the GW upper limit using data selected in the first part.
\subsection{Data Selection}
To perform this analysis, one needs to download the data of the LIGO detectors at both Hanford and Livingston from \href{https://www.gw-openscience.org/data/}{O2 data release} (\url{https://www.gw-openscience.org/data/})~\cite{LIGOScientific:2019vic} with a 16 kHz sampling frequency.
Each file covers a 4096 s period of measurement, and the file name contains the start time of the measurement, which is referred to as the ``GPS start time''. 
In  each  file,  there  are  in  general  two  types  of data – the strain time series $h(t_i)$ and some auxiliary data such as the data quality (DQ) mask label associated with each strain measurement.
The DQ mask is a 7-bit binary number each of which indicates whether a certain type of check is passed (value=1) or not (value=0).
We convert this binary number into a decimal digit.  
For example, there is no data at time $t_i$ if DQ$(t_i) = (0000000)_{10}= 0$.
On the contrary, data is present if this value is nonzero.

Following the LIGO stochastic gravitational wave analysis \cite{TheLIGOScientific:2016dpb,LIGOScientific:2019vic}, we first downsample the 16 kHz strain data to 4 kHz.
We then select out the timestamps at which both detectors are taking data properly, \ie~the times $t_i$ at which both $\text{DQ}_H(t_i)\neq 0$ and $\text{DQ}_L(t_i)\neq 0$.
After doing that we get, for each data file, a list of GPS times at which data in both detectors is available.
We then combine the lists of GPS times within a file and across neighbouring files into continuous segments.
The segments whose duration $\geq 600$ s are further picked out to perform a stationarity cut following Ref. \cite{Abbott:2009ws,TheLIGOScientific:2016dpb}.
When we perform the stationarity cut, we also notch out frequencies at which
the data exhibits narrowband coherent lines that are known to be instrumental or environmental artifacts~\cite{abbott2016characterization,thrane2013correlated,thrane2014correlated,covas2018identification}.
The list of the notched frequency bands we used can be found on the public data release page: \url{https://dcc.ligo.org/LIGO-T1900058/public}.

\subsection{Analysis}
After selecting a clean list of strain data and GPS time which satisfies the requirements on data quality and stationarity, we then use the cross-correlation method to estimate the SGWB signal. The spectrum of the  GW background is estimated with the cross-correlation statistic $\hat{C}(f)$ \cite{LIGOScientific:2019vic}, defined as
\beqn
\hat{C}(f) &\equiv& \frac{2}{T}\frac{\textrm{Re}[\tilde{s}^*_1(f)\tilde{s}_2(f)]}{\gamma_T(f)S_0(f)},\\
\langle \hat{C}(f) \rangle &=& \Omegagw(f) ,
\eeqn
where $S_0 = 3H_0^2/(10\pi^2 f^3)$, $H_0$ is the Hubble parameter, $\tilde{s}_{1,2}(f)$ are the Fourier transfroms of the strain data of both detectors, $T = 192$ s is the segment duration of the Fourier transforms, and $\langle...\rangle$ indicates the average over all such 192s-segments. In the limit that the GW signal is negligible comparing to the instrumental noise, the variance of  $\hat{C}(f)$ is given by
\beqn
\sigma^2(f) = \frac{1}{2T\Delta f}\frac{\bar{P}_1(f)\bar{P}_2(f)}{\gamma_T^2(f)S_0^2(f)} ,
\eeqn
where $\Delta f = 1/32$ Hz, and $\bar{P}_{1,2}(f)$ are the one-sided power spectrum of each detector, which are  obtained as an average over two neighbouring segments~\footnotemark[1], 
\beq
\bar P_{i,I}\equiv \frac{P_{i,I-1}+P_{i,I+1}}{2},\label{eq:P_avg}
\eeq
in which the subscript $i$ and $I$ labels the detector and the 192s-segments, respectively. For each 192s-segment, we use the broadband estimator for any spectral shape of the gravitational wave background,
\beqn
\hat{\Omega}_{\rm ref} &\equiv& \frac{\sum_k \omega(f_k)^{-2}\hat{C}(f_k)\sigma^{-2}(f_k)}{\sum_k \omega(f_k)^{-1} \sigma^{-2}(f_k)}\\
 \langle \hat{\Omega}_{\rm ref} \rangle &=& \Omegagw(\fref)
\eeqn
where the weight function $\omega(f) \equiv \Omegagw(f_{\rm ref})/\Omegagw(f)$, $f_k$ are discrete frequencies between 20 and 1726 Hz with the interval of $1/32$ Hz. The uncertainty of the optimal estimator is,
\beq
\sigma^{-2}_{\Omega} = \sum_k \omega(f_k)^{-1} \sigma^{-2}(f_k).\\
\eeq
After calculating $\hat{\Omega}_{\rm ref}$ and $\sigma^2_{\Omega}$ for all 192s-segments, the ensemble average over all the segments is obtained from
\beqn
\sigma^2_{\Omega,{\rm tot}}&=&\frac{1}{\sum_I \sigma_{\Omega,I}^{-2}}\,,\\
\mu &=& \frac{\sum_I {\hat{\Omega}_{{\rm ref},I}}{\sigma_{\Omega,I}^{-2}}}{\sum_I \sigma_{\Omega,I}^{-2}}\,.
\eeqn
The signal-to-noise ratio (SNR) can be calculated by $\mathrm{SNR}=\mu/\sigma$, where $\sigma\equiv\sqrt{\sigma_{\Omega,{\rm tot}}^2}$.
In the absence of detection signal, we  set the 95\% confidence level upper limit by
\beq
\Omegaref^{\rm up.~lim.}=\Omegagw^{\rm up. ~lim}(\fref)=2\times\sigma_{\Omega,{\rm tot}}(\fref)\,.
\eeq
Note that this upper limit depends on the choice of the reference frequency $\fref$. In the calculation of Fourier transforms $\tilde{s_i}(f)$ and power spectral density $P_i(f)$, we use the 50\% overlapping Hann windows to avoid spectral leakage~\cite{TheLIGOScientific:2016dpb,LIGOScientific:2019vic,Abbott:2021xxi}. 
\footnotetext[1]{This means we need at least 3 continuous 192s-segments to obtain a power spectrum. 
Those 192s-segments without a neighbor on both sides do not have an associated power spectrum.}
\bibliography{ref}
\end{document}